\documentclass{aa}  
\usepackage{graphicx}
\usepackage{txfonts}
\begin{document}

\title{HD\,152384: A 5 Myr old debris disk system showing a gaseous ring of refractory elements at the dust sublimation radius\thanks{Based on observations collected at the European Southern Observatory under ESO programmes 109.23HY.001 and 109.23K8.001.}}
\titlerunning{A young debris disk system showing a gaseous ring of refractory elements}
   \author{M. E. van den Ancker\inst{1}
          \and
          D. P. Iglesias\inst{2,3}
          \and 
          O. Pani\'c\inst{2}
          \and
          M. G. Petr-Gotzens\inst{1}
          \and
          M. Vioque\inst{1}
          }

   \institute{European Southern Observatory, Karl-Schwarzschild-Str. 2, D--85748 Garching bei M\"unchen, Germany\\
             \email{mvandena@eso.org}
             \and
             School of Physics and Astronomy, University of Leeds, Woodhouse Lane, Leeds LS2 9JT, UK
             \and
             School of Mathematical and Physical Sciences, University of Sheffield, Hounsfield Road, Sheffield S3 7RH, UK
            }

   \date{Received 17 April 2026; Accepted 13 October 2026}

  \abstract
  {Circumstellar gas around young stars provides key insights into planet formation and disk evolution. HD\,152384, a young (5 Myr) A0\,V star in Sco-Cen, exhibits hybrid disk characteristics with a modest infrared excess due to small dust grains and atomic emission lines dominated by refractory elements, but the disk's structure and origin remain unclear.}
  {We aim to characterise the kinematics and excitation of the circumstellar gas around HD\,152384, and to constrain the physical properties of the emitting region.}
  {We use new optical and near-infrared high spectral resolution data from the UVES and CRIRES instruments at the VLT to revisit the circumstellar environment of HD\,152384. Earlier studies suggested the presence of a broad, gas-rich circumstellar disk; here, we re-examine this scenario.}
  {The new observations reveal a rich emission-line spectrum, consisting of 74 narrow double-peaked emission lines of iron, magnesium, silicon and calcium. All lines share similar profiles, with 103~km\,s$^{-1}$ peak separation. Modelling of gas in Keplerian rotation indicates that the gas is confined to a geometrically thin ring at a radial distance of 0.09--0.12 au from the central star -- close to the dust sublimation radius (0.10--0.14~au, assuming a dust sublimation temperature of $\sim$1500~K). The detected Fe\,{\sc i} emission comes from an optically thick region. However, the Mg\,{\sc i} lines are optically thin, allowing us to derive a mass of 2.2 $\times$ 10$^{22}$~kg (or 3.6 $\times$ 10$^{-3}$~M$_\oplus$) in magnesium gas confined to the geometrically thin ring.}
  {The inferred mass and composition of the circumstellar dust and gas in the HD\,152384 system are consistent with a secondary origin related to rocky planetesimal collisions. We suggest that the atomic gas seen in our observations is due to the sublimation of small dust grains created in such collisions, and spiralling inward to the dust sublimation radius, creating an accumulation of atomic gas at this location. We conclude that HD\,152384 represents a young analogue of gas-bearing debris disk systems such as $\beta$~Pictoris and of the gaseous debris disks around metal-polluted white dwarfs (in which a similar phenomenology is observed), offering a new window into the early violent phase of rocky planetesimal processing.}

   \maketitle
   \nolinenumbers
%

\section{Introduction}
Circumstellar gas is a key tracer of planet formation and disk evolution processes around young stars. Disks around young A-type stars display a variety of morphologies and chemical signatures, ranging from volatile-rich primordial gas seen in classical Herbig Ae/Be and transition disks to debris disks -- systems hosting extrasolar Edgeworth-Kuiper belts in which the gas content may be secondary \citep[e.g.][]{Wyatt2008,Hughes2018,Waters2015,vanderMarel2023}. This includes systems in which the gas appears linked to the breakup of rocky material, with asymmetries and volatile signatures consistent with recent or localised collisions \citep[e.g.][]{Kiefer2014,Rebollido2018}. This diversity also encompasses transition disks, which occupy an intermediate evolutionary stage characterised by partially cleared inner regions and on-going gas dispersal, often exhibiting a mix of primordial and secondary gas signatures that reflect both disk evolution and potential planet–disk interactions \citep[e.g.][]{Marino2020,Nakatani2021,Iglesias2025}.

\object{HD\,152384} is a bright ($G$ = 7.01), young \citep[Age $\sim$ 5 Myr; ][]{Iglesias2023} A0\,V star in the Sco-Cen association \citep[][]{vandenAncker2021}. Based on its 12~$\mu$m infrared excess and above the main-sequence position in the HR-diagram, \cite{Iglesias2023} classified HD\,152384 as a 'hybrid' disk candidate, at an intermediate stage of evolution between protoplanetary accretion disks and debris disks with no indication of on-going accretion. The optical spectrum of the star shows double-peaked emission from iron, magnesium, silicon and calcium, but pure absorption lines in hydrogen and helium. This implies the presence of circumstellar material highly depleted in volatile elements, potentially indicating the presence of an evolved or collisionally processed disk \citep{vandenAncker2021}.

The Gaia Data Release 3 \citep{Gaia2023} shows a close companion to HD\,152384, with an angular separation of 1.60\arcsec\ corresponding to 205 au at 128~pc. The companion is significantly fainter ($G$ = 13.7) than HD\,152384, but the Gaia Data Release 3 Catalogue does not include a parallax or proper motion measurement for this companion star. We note that an archival VLT/ISAAC \citep{Moorwood1998} K$_s$-band image of HD\,152384 taken on 9 June 2010, shows the same point-source companion detected by Gaia, at a magnitude difference of $\Delta$K$_s$ = 3.06. The resulting $G-K_s$ colour of 3.9 is compatible with a late K- or early M-type star. It is at present unclear whether HD\,152384 is a physical binary, or if this apparent companion is just seen projected on the sky close to the bright A-type star. Based on an anomaly between the Hipparcos and Gaia DR3 proper motions of the bright primary star, \cite{Kervella2022} also classified HD\,152384 as a suspected binary system. However, it is unlikely that the wide companion noted above could be responsible for this proper motion anomaly. If confirmed, the proper motion anomaly would indicate the presence of a closer companion.

In this paper, we report new high-resolution spectroscopic observations of HD\,152384 with CRIRES (covering the near-IR) and UVES (covering optical wavelengths), providing a fresh look at the structure and composition of the gas in this enigmatic object.

\section{Observations and data reduction}
%
   \begin{figure*}
   \centering
   \includegraphics[width=16.0cm]{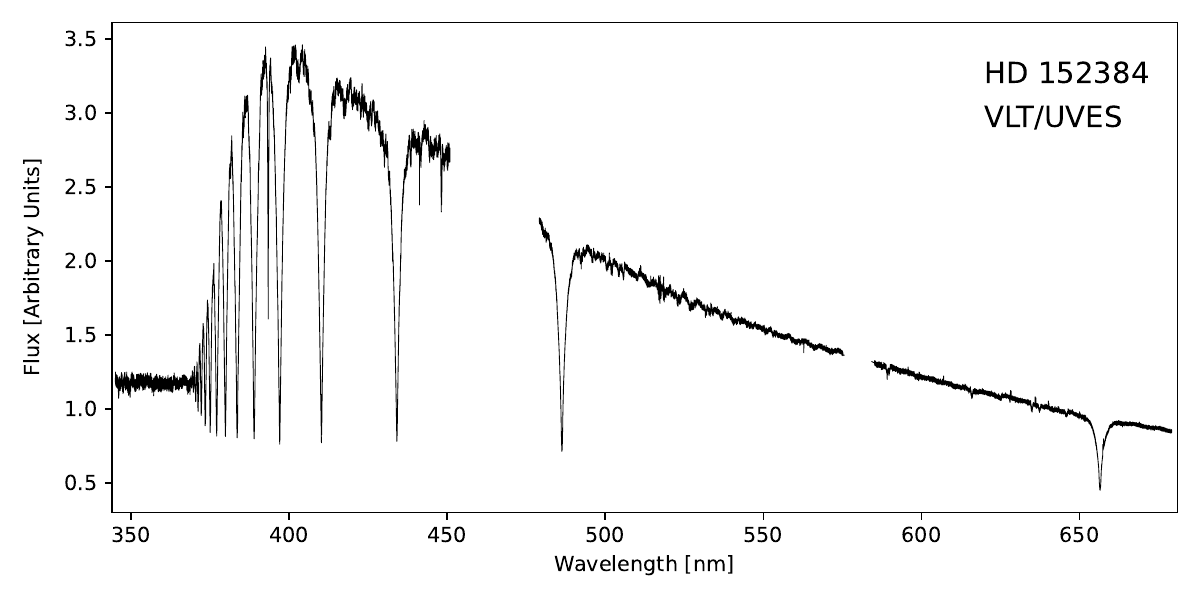}
   \caption{Full UVES reference spectrum of HD\,152384. Shown are the spectra of the blue arm (329.5--451.0 nm) and the red arm (479.1--679.2 nm). The red arm spectrum has a small gap from 575.6 to 584.3 nm due to the gap between the two chips in the red arm detector mosaic.}
              \label{Fig0}
    \end{figure*}
\begin{table*}
  \caption[]{Observing log of high-resolution spectroscopic observations of HD\,152384.}
  \label{Table1}
\tabcolsep0.17cm
\vspace*{-0.4cm}
\begin{flushleft}
\begin{tabular}{lcccccccc}
\hline\noalign{\smallskip}
 INS       &   Prog. ID     &     MJD         &   Obs. Date \& Time (UT)    &     Wavelength range [nm] &  $T_{\rm exp}$ [s] & Airmass & Seeing ["]\\ 
\noalign{\smallskip}
\hline\noalign{\smallskip}
 CRIRES    &   109.23HY.001 &	59708.29099 &  2022-05-09   ~~06:59:01.2	&		~949.6--1120.5	& 300.0	 & 1.02 & 2.1\\
 CRIRES    &   109.23HY.001 &   59708.29584 &  2022-05-09   ~~07:06:00.3        &		1118.3--1357.5	& 300.0	 & 1.02 & 2.3\\
 CRIRES    &   109.23HY.001 &   59708.30039 &  2022-05-09   ~~07:12:33.8	&		1461.3--1828.4	& 300.0	 & 1.03 & 2.3\\
 CRIRES    &   109.23HY.001 &   59708.30508 &  2022-05-09   ~~07:19:18.8	&		1445.8--1796.3	& 300.0	 & 1.03 & 2.0\\
 CRIRES    &   109.23HY.001 &   59708.30983 &  2022-05-09   ~~07:26:09.2	&		1972.3--2452.2	& 300.0	 & 1.04 & 1.9\\
 CRIRES    &   109.23HY.001 &   59708.31457 &  2022-05-09   ~~07:32:58.7	&		1945.9--2501.0	& 300.0	 & 1.04 & 1.8\\
 \\                                                                                                                             
 UVES	   &   109.23K8.001 &   59774.05724 &  2022-07-14   ~~01:22:25.8	&	         479.1--679.2    & 	45.0	 &1.03 & 1.8\\
 UVES	   &   109.23K8.001 &   59774.05729 &  2022-07-14   ~~01:22:29.7	&	         329.5--451.0    & 	45.0	 &1.03 & 1.8\\
 UVES	   &   109.23K8.001 &   59783.12980 &  2022-07-23   ~~03:06:54.5	&	         479.1--679.2    & 	45.0	 &1.08 & 0.6\\
 UVES	   &   109.23K8.001 &   59783.12984 &  2022-07-23   ~~03:06:58.4	&	         329.5--451.0    & 	45.0	 &1.08 & 0.6\\
 UVES	   &   109.23K8.001 &   59785.09398 &  2022-07-25   ~~02:15:19.6	&	         479.1--679.2    & 	45.0	 &1.03 & 1.5\\
 UVES	   &   109.23K8.001 &	59785.09402 &  2022-07-25   ~~02:15:23.1	&	         329.5--451.0    &      45.0	 &1.03 & 1.5\\
\noalign{\smallskip}
\hline
\end{tabular}
\end{flushleft}
\end{table*}
%
%
   \begin{figure*}
   \centering
   \includegraphics[height=21.5cm]{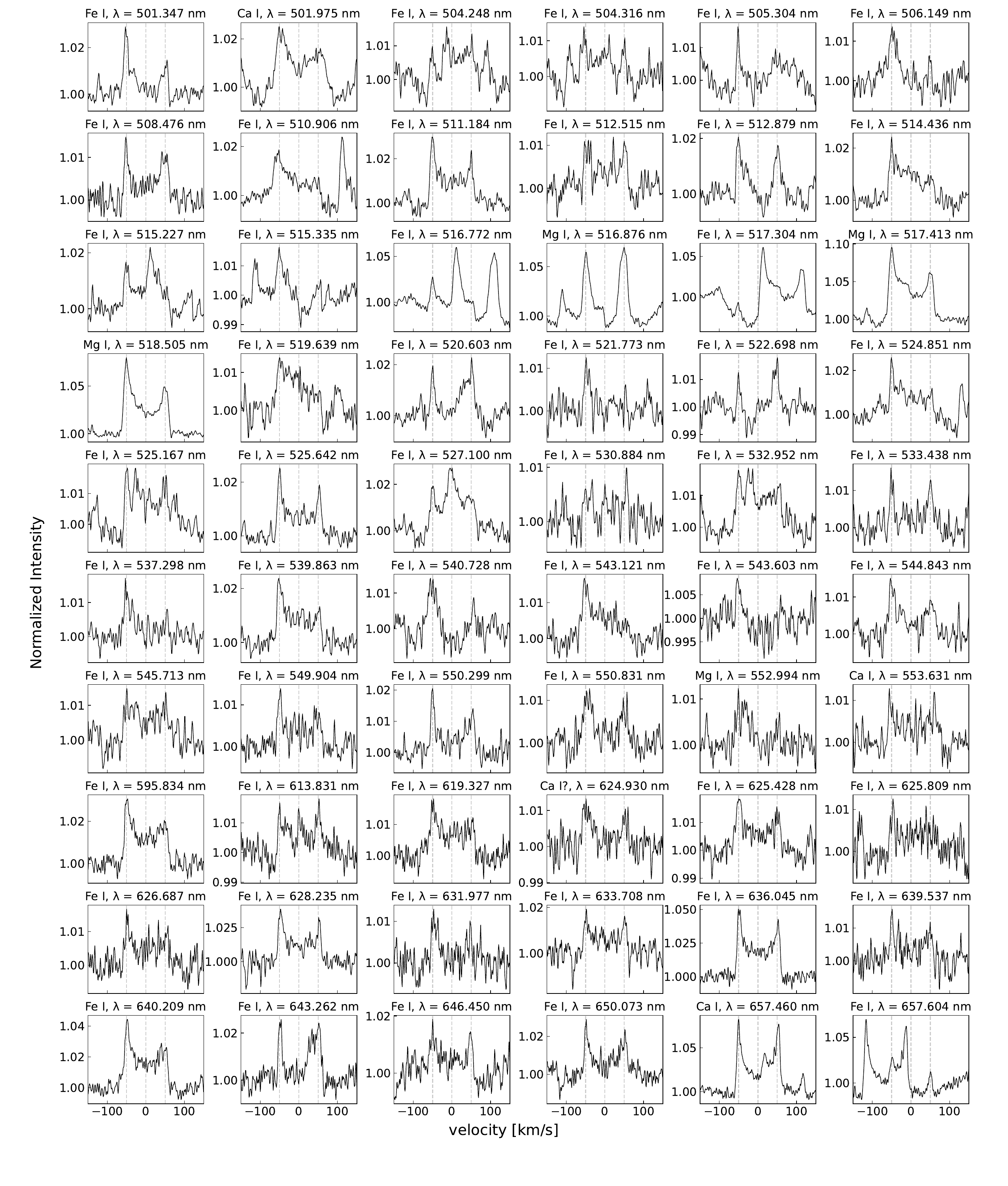}
   \vspace*{-0.9cm}
   \caption{Detected emission lines in the UVES spectrum of HD\,152384. Velocity scales for all lines have been corrected for the systemic radial velocity. In each panel, the continuum flux at velocity 0 has been normalised to unity by linear interpolation of the spectrum between the velocity ranges $-150$ to $-100$ km\,s$^{-1}$ and $+100$ to $+150$ km\,s$^{-1}$. Also shown (dashed vertical lines) are the systemic velocity (at 0 km\,s$^{-1}$), as well as the expected locations of the emission profile peaks at $-51.5$ and $+51.5$ km\,s$^{-1}$.}
              \label{Fig1}
    \end{figure*}
   \begin{figure*}
   \centering
   \includegraphics[width=18.5cm]{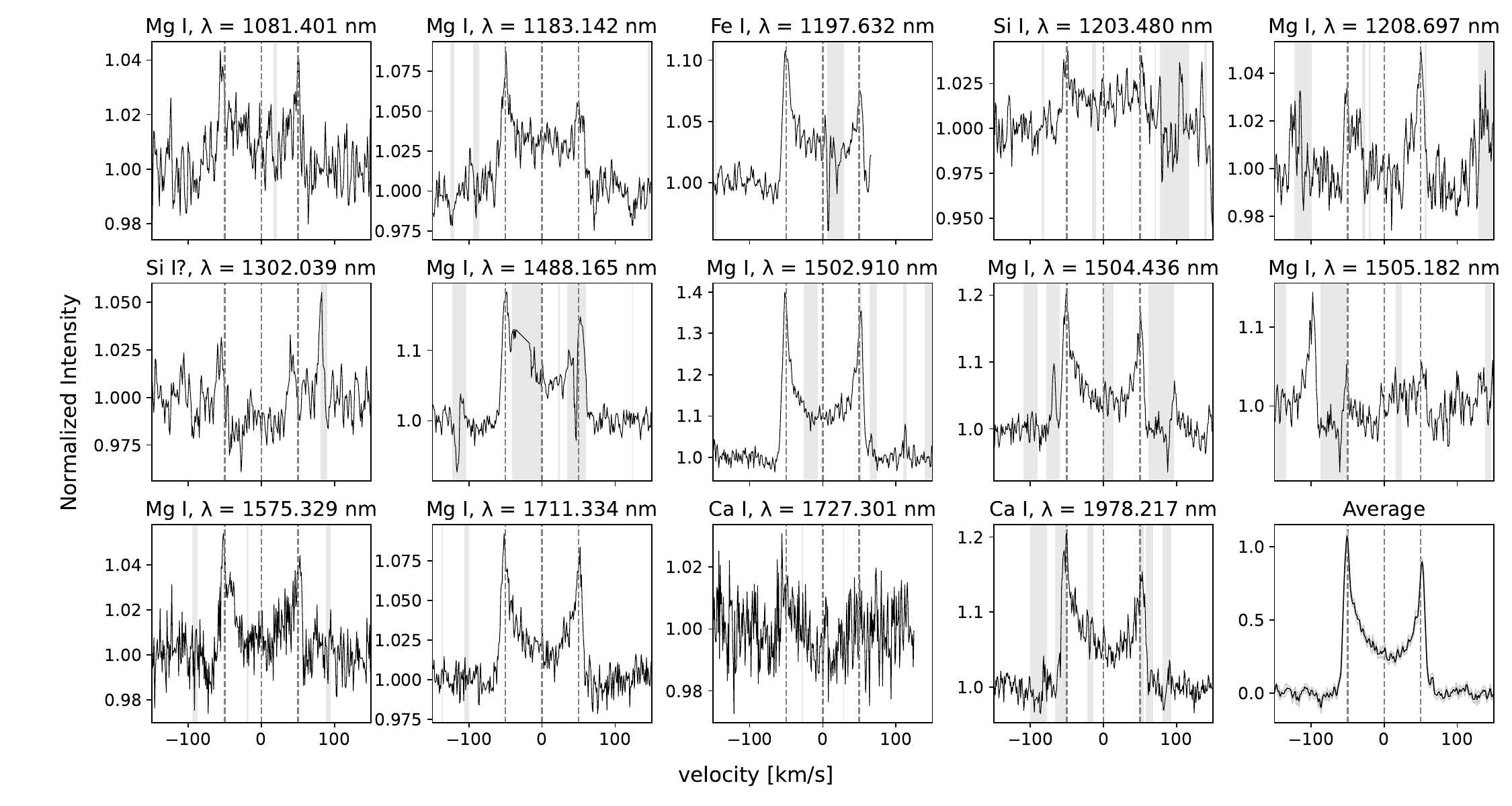}
   \caption{Detected emission lines in the CRIRES spectrum of HD\,152384. Velocity scales for all lines have been corrected for the systemic radial velocity. The dashed vertical lines show the systemic velocity (at 0 km\,s$^{-1}$), as well as the expected locations of the emission profile peaks at $-51.5$ and $+51.5$ km\,s$^{-1}$. Regions of the spectrum affected by strong telluric absorption are marked by a light grey background. In each panel, the continuum flux at velocity 0 has been normalised to unity by linear interpolation of the spectrum between the velocity ranges $-150$ to $-100$ km\,s$^{-1}$ and $+100$ to $+150$ km\,s$^{-1}$. Also shown (bottom right panel) is the average continuum-subtracted CRIRES line profile, normalised to peak intensity one. The grey lines in this panel illustrate the standard deviation in the average line profile.}
              \label{Fig2}
    \end{figure*}
New high spectral resolution optical spectra of HD\,152384 were obtained with UVES \citep[Ultraviolet and Visual Echelle Spectrograph;][]{Dekker2000} on the VLT on 14, 23 and 25 July 2022 (Table~\ref{Table1}). UVES is a two-arm cross-dispersed spectrograph with flexible wavelength coverage, where the blue and red arms can simultaneously cover the ranges 300--500 nm and 420--1100 nm, respectively. Our observations were performed using the dichroic mode with central wavelengths 390+580 nm, and the 0.5\arcsec\ slit width for both arms, resulting in spectral resolutions of 65,000 and 74,500 for the blue and red arms, respectively. Exposure times were 45 seconds for both the red and blue arms, achieving a S/N of $>$ 100 across the wavelength range. A position angle of 160 degrees on the sky was selected for the slit. This position angle is perpendicular to the position angle of the close companion, minimising any stray light due to this star. Given the good seeing conditions under which some of the data were taken (e.g. on 2022-07-23; Table~\ref{Table1}), any contamination by light from the nearby companion is expected to be minimal in these data.

The UVES data were reduced using the UVES data reduction pipeline version 6.4.1\footnote{\url{http://www.eso.org/sci/software/pipelines/}}, after which the spectra were corrected for telluric absorption using a synthetic transmission spectrum computed using the molecfit package\footnote{\url{http://www.eso.org/sci/software/pipelines/skytools/molecfit}}, fitted to the red arm data. We note that molecfit updates the errors associated with the spectra to correctly propagate the additional noise contribution in spectral regions affected by strong telluric absorption lines. As a final step in the reduction of the individual spectra, the wavelength calibration of the spectra was corrected for the barycentric radial velocity. Within errors, the individual spectra taken on different dates (and under different seeing conditions) appear identical, providing further evidence that the resulting spectra are not contaminated by emission from the possible nearby companion. A cross-correlation function analysis of the stellar absorption features seen in the individual spectra gives an upper limit to possible radial velocity variations in the time period covered by our observations of 1.6~km\,s$^{-1}$. In order to improve the total S/N, we also created a 'reference spectrum' by computing the statistical median of the individual spectra -- previously normalised to the continuum to have the same flux level before combining.  The full resulting UVES reference spectrum is shown in Fig.~\ref{Fig0} and cut-outs of this spectrum around spectral regions of interest are shown in Fig.~\ref{Fig1}.

New high spectral resolution infrared spectra of HD\,152384 covering the Y, J, H and K-bands (using the Y1029, J1228, H1559, H1582, K2147, K2192 settings) were obtained with the CRIRES spectrograph on the VLT \citep{Dorn2023} on 9 May 2022. Full details of these observations are given in Table~\ref{Table1}. For each wavelength setting a single AB nodding cycle was obtained, with a total integration time of 300 seconds per setting, resulting in a S/N $>$ 100 for the parts of the spectrum with good atmospheric transmission. The spectral resolution $\lambda/\Delta \lambda$ in these spectra (taken through a 0.2\arcsec\ wide slit) was measured from narrow telluric absorption lines. It was found to vary between $R$ = 102,000 and $R$ = 107,000 between the spectra taken in the different wavelength settings. The slit was oriented along the parallactic angle (varying between $38.1^\circ$ and $64.0^\circ$) to minimise slit losses caused by differential atmospheric refraction. Although these data were taken under relatively poor seeing conditions in the optical, the measured infrared image quality (the FWHM in the spatial direction of a Gaussian fitted to the 2D spectral trace) in the CRIRES data ranges from 1.5\arcsec\ (Y-band) to 1.1\arcsec\ (K-band), ensuring sufficient spatial separation between the primary and possible companion. The CRIRES data were reduced using the CRIRES data reduction pipeline version 1.6.7\footnote{\url{http://www.eso.org/sci/software/pipelines/}} with default parameters. After correcting for flatfielding and detector non-linearity, the data reduction pipeline subtracts the images for the A and B beams of the nodding pattern, and then extracts and assigns an individual wavelength calibration to the signal in the A and B beams in order to correct for the variation of wavelength calibration with position in the slit. Finally, the CRIRES pipeline produces a combined data product with the sum of the A and B beam spectra. A  known issue with the employed version of the CRIRES data reduction pipeline is that in certain cases the method used to estimate of the uncertainties (which includes error contributions from shot noise, detector read-out-noise, bad pixels, correction for cosmic rays, detector non-linearity, error terms due to flatfielding, and the noise introduced by the spectral extraction) underestimates the scatter in the data. Therefore we re-scaled this error spectrum produced by the pipeline to match the rms scatter measured in line-free continuum regions in the spectrum. As for UVES, the final CRIRES spectra were subsequently corrected for telluric absorption using a synthetic transmission spectrum computed using the molecfit package, and corrected for the barycentric radial velocity. Excerpts of the CRIRES spectra covering wavelengths of interest are shown in Fig.~\ref{Fig2}.

The UVES and CRIRES spectra were not obtained under photometric conditions. In addition, the significant slit losses (which are not corrected for by the data reduction pipeline) are poorly characterised for both data-sets. Therefore it is not possible to obtain a reliable absolute flux calibration for these spectra, and we will only analyse the line strengths relative to the stellar continuum.

\section{Results and analysis}
\subsection{Line identifications}
The new UVES and CRIRES spectra display a rich spectrum of narrow double-peaked emission lines superimposed on the strong stellar continuum. In the UVES spectra emission lines are only found longward of 500 nm. However, we note that shortward of 500 nm the stellar continuum of the A-type star becomes stronger and due to the higher photon noise, weaker lines could not be detected against the strong background emission. A total of 60 emission lines are detected in the UVES spectra and an additional 14 emission lines were found in the CRIRES spectra. They are shown in Figures~\ref{Fig1} and \ref{Fig2}. Line positions and intensities are given in Tables~\ref{Table2} and ~\ref{Table3}.
   
   We identify the spectral lines using the line list by \cite{vanHoof2018}\footnote{\url{https://linelist.pa.uky.edu/newpage/}}. For the line identifications, we started with the strongest detected lines, and then searched in the spectral line list for lines from the same multiplet to identify the weaker lines. This allowed us to identify most lines. For the remaining weaker lines, we selected the most plausible line identification from the list. These are identified with a question mark in Tables~\ref{Table2} and \ref{Table3}.

   The resulting optical-to-near-IR emission lines found in the spectrum of HD\,152384 are dominated by lines from Fe\,{\sc i} (53 lines) and Mg\,{\sc i} (13 lines). There are also clear detections of strong lines from Si\,{\sc i} (two lines) and Ca\,{\sc i} (six detected lines). As was already noted in the lower-resolution X-Shooter spectrum reported by \citet{vandenAncker2021}, emission lines from H\,{\sc i} and He\,{\sc i}, as well as UV excess due to accretion, are conspicuously absent, consistent with the presence of circumstellar gas which is rich in refractory elements, but depleted in volatile species. We note, however, that excitation effects and the different wavelength regimes in which lines of volatiles and refractory species may manifest, may also explain the non-detection of volatile species in our data.

   Equivalent widths (EWs) for all detected lines were measured using the {\tt specutils} {\tt equivalent\_width} routine from the {\tt astropy} library \citep{Astropy2013, Astropy2018, Astropy2022}. In all cases EWs were integrated in the spectral region from $-60$ to $+60$ km\,s$^{-1}$. Prior to integration, the continuum level in the spectral region around the line was normalised to unity by linear interpolation of the spectrum in the velocity ranges $-150$ to $-100$ km\,s$^{-1}$ and $+100$ to $+150$ km\,s$^{-1}$. All measured EWs, as well as their errors, are reported in Tables~\ref{Table2} and \ref{Table3}. The error estimates on the EW values given in these Tables includes the errors due to photon noise, the error due to the determination of the continuum level, as well as the uncertainty introduced by the telluric correction. For some lines, the line profiles of closely spaced lines overlap. In this case these lines are marked as ``blend'' in the Tables and the reported equivalent width (EW) is the sum of both lines.

\subsection{Kinematics}
   \begin{figure*}
   \centering
   \includegraphics[width=16.0cm]{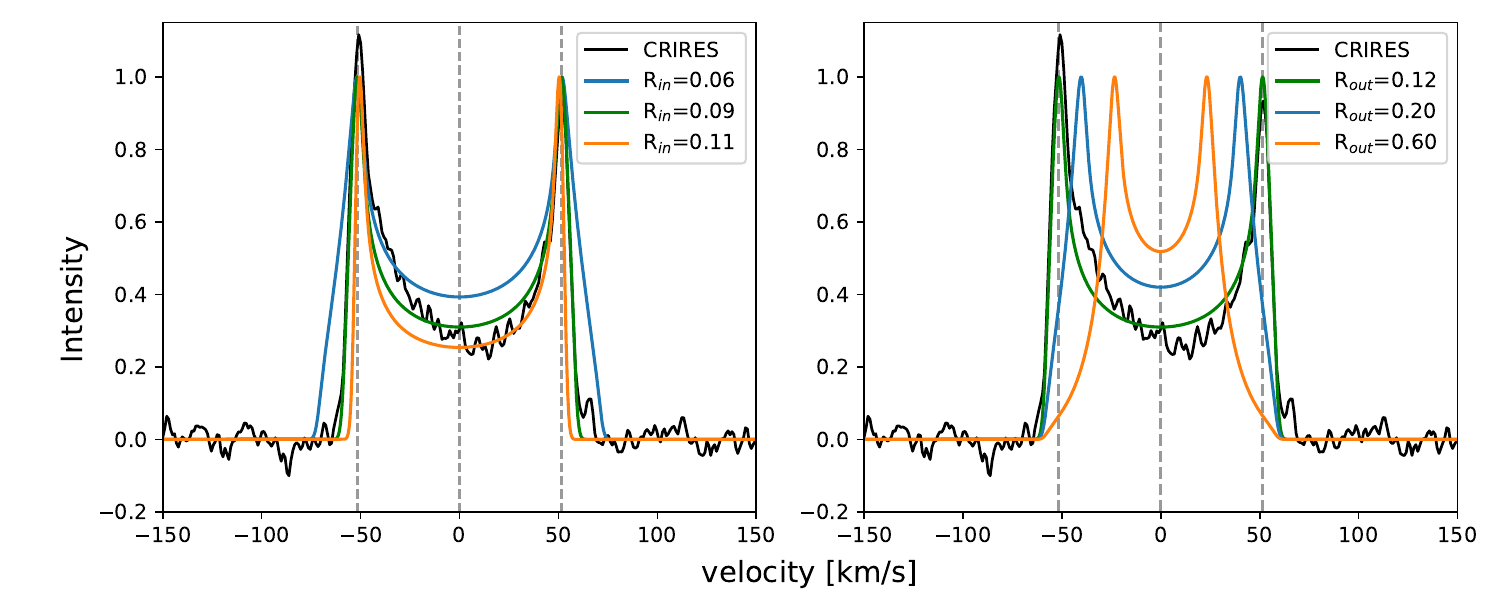}
   \caption{Simulated emission lines from a flat circumstellar disk using the toy disk model. The black lines show the continuum-subtracted average line profile of the CRIRES spectra of HD\,152384, whereas the green lines show the model with the parameters $i = 24^\circ$ (corresponding to the stellar inclination), $\alpha = 1.0$, $R_{\rm in} = 0.09$ au, $R_{\rm out}$ = 0.12 au (corresponding to the best-fit parameters). In each panel the orange and blue lines illustrate the effect of varying one model parameter (left: inner radius in au, right: outer radius in au).}
              \label{Fig3}
    \end{figure*}
\begin{table}
  \caption[]{Detected emission lines in the UVES spectra of HD\,152384.}
  \label{Table2}
 \tabcolsep0.14cm
 \vspace*{-0.4cm}
\begin{center}
\begin{tabular}{clccc}
\hline\noalign{\smallskip}
$\lambda_{\rm lab.}$ [nm] & Ident. & Multiplet & Trans. & EW [m\AA]\\
\noalign{\smallskip}
\hline\noalign{\smallskip}
501.347  & Fe\,{\sc i}     & $a^5\mathrm{F}-z^5\mathrm{F}^o$      &  5--5  &  $-15.8~\pm~0.8$\\
501.979  & Ca\,{\sc i}     & $^1\mathrm{D}-^3\mathrm{P}^o$        &  2--1  &  $-24.9~\pm~1.8$\\
504.248  & Fe\,{\sc i}     & $a^5\mathrm{F}-z^5\mathrm{F}^o$      &  3--2  &  $-3.8~\pm~1.3$ (blend)\\
504.316  & Fe\,{\sc i}     & $a^3\mathrm{F}-z^3\mathrm{F}^o$      &  4--3  &  $-3.8~\pm~1.3$ (blend)\\
505.304  & Fe\,{\sc i}     & $a^5\mathrm{F}-z^5\mathrm{F}^o$      &  4--4  &  ~$-7.4~\pm~0.5$\\
506.149  & Fe\,{\sc i}     & $a^5\mathrm{D}-z^7\mathrm{D}^o$      &  4--3  &  $-13.4~\pm~0.9$\\

508.476  & Fe\,{\sc i}     & $a^5\mathrm{F}-z^5\mathrm{F}^o$      &  3--3  &  $-12.1~\pm~0.7$\\    
510.906  & Fe\,{\sc i}     & $a^3\mathrm{F}-z^3\mathrm{F}^o$      &  3--2  &  $-12.0~\pm~1.2$\\
511.184  & Fe\,{\sc i}     & $a^5\mathrm{D}-z^7\mathrm{D}^o$      &  4--4  &  $-24.5~\pm~0.6$\\
512.515  & Fe\,{\sc i}     & $a^5\mathrm{F}-z^5\mathrm{F}^o$      &  1--1  &  ~$-9.3~\pm~0.6$\\
512.879  & Fe\,{\sc i}     & $a^5\mathrm{F}-z^5\mathrm{F}^o$      &  4--5  &  $-12.8~\pm~0.9$\\
514.436  & Fe\,{\sc i}     & $a^5\mathrm{F}-z^5\mathrm{F}^o$      &  3--4  &  $-21.2~\pm~0.7$\\

515.227  & Fe\,{\sc i}     & $a^5\mathrm{F}-z^5\mathrm{F}^o$      &  2--3  & $-20.5~\pm~0.8$ (blend)\\
515.335  & Fe\,{\sc i}     & $a^5\mathrm{F}-z^5\mathrm{F}^o$      &  1--2  & $-20.5~\pm~0.8$ (blend)\\
516.772  & Fe\,{\sc i}     & $a^5\mathrm{D}-z^7\mathrm{D}^o$      &  4--5  & $-57.2~\pm~4.3$ (blend)\\
516.876  & Mg\,{\sc i}     & $^3\mathrm{P}^o-^3\mathrm{S}$        &  0--1  & $-57.2~\pm~4.3$ (blend)\\
517.304  & Fe\,{\sc i}     & $a^3\mathrm{F}-z^3\mathrm{F}^o$      &  4--4  & $-97.9~\pm~3.0$ (blend)\\
517.413  & Mg\,{\sc i}     & $^3\mathrm{P}^o-^3\mathrm{S}$        &  1--1  & $-97.9~\pm~3.0$ (blend)\\

518.505  & Mg\,{\sc i}     & $^3\mathrm{P}^o-^3\mathrm{S}$        &  2--1  & $-61.4~\pm~1.2$\\
519.639  & Fe\,{\sc i}     & $a^3\mathrm{F}-z^3\mathrm{F}^o$      &  3--3  & $-12.9~\pm~1.1$\\
520.603  & Fe\,{\sc i}     & $a^5\mathrm{D}-z^7\mathrm{D}^o$      &  2--2  & $-17.3~\pm~1.2$\\     
521.773  & Fe\,{\sc i}     & $a^3\mathrm{F}-z^3\mathrm{F}^o$      &  2--2  & ~$-3.3~\pm~0.6$\\
522.698  & Fe\,{\sc i}     & $a^5\mathrm{D}-z^7\mathrm{D}^o$      &  1--1  & ~$-4.2~\pm~0.9$\\     
524.851  & Fe\,{\sc i}     & $a^5\mathrm{D}-z^7\mathrm{D}^o$      &  2--3  & $-15.2~\pm~0.7$\\    

525.167  & Fe\,{\sc i}     & $a^5\mathrm{D}-z^7\mathrm{D}^o$      &  0--1  & $-18.8~\pm~1.0$\\
525.642  & Fe\,{\sc i}     & $a^5\mathrm{D}-z^7\mathrm{D}^o$      &  1--2  & $-18.5~\pm~0.7$\\
527.100  & Fe\,{\sc i}     & $a^5\mathrm{F}-z^5\mathrm{D}^o$      &  5--4  & $-28.3~\pm~0.9$\\
530.884  & Fe\,{\sc i}     & $a^3\mathrm{F}-z^3\mathrm{F}^o$      &  2--3  & ~$-4.5~\pm~0.8$\\  
532.952  & Fe\,{\sc i}     & $a^5\mathrm{F}-z^5\mathrm{D}^o$      &  4--3  & $-20.1~\pm~0.8$\\
533.438  & Fe\,{\sc i}     & $a^3\mathrm{F}-z^3\mathrm{F}^o$      &  3--4  & $-19.8~\pm~1.2$\\

537.298  & Fe\,{\sc i}     & $a^5\mathrm{F}-z^5\mathrm{D}^o$      &  3--2  & ~$-6.9~\pm~0.7$\\
539.863  & Fe\,{\sc i}     & $a^5\mathrm{F}-z^5\mathrm{D}^o$      &  4--4  & $-20.3~\pm~0.8$\\
540.728  & Fe\,{\sc i}     & $a^5\mathrm{F}-z^5\mathrm{D}^o$      &  2--1  & ~$-3.9~\pm~1.1$\\
543.121  & Fe\,{\sc i}     & $a^5\mathrm{F}-z^5\mathrm{D}^o$      &  3--3  & $-14.3~\pm~0.9$\\
543.603  & Fe\,{\sc i}     & $a^5\mathrm{F}-z^5\mathrm{D}^o$      &  1--0  & ~$-2.5~\pm~0.7$\\
544.843  & Fe\,{\sc i}     & $a^5\mathrm{F}-z^5\mathrm{D}^o$      &  2--2  & ~$-9.0~\pm~0.9$\\

545.713  & Fe\,{\sc i}     & $a^5\mathrm{F}-z^5\mathrm{D}^o$      &  1--1  & $-11.0~\pm~0.9$\\
549.904  & Fe\,{\sc i}     & $a^5\mathrm{F}-z^5\mathrm{D}^o$      &  1--2  & $-11.3~\pm~0.6$\\
550.299  & Fe\,{\sc i}     & $a^5\mathrm{F}-z^5\mathrm{D}^o$      &  3--4  & $-13.3~\pm~0.7$\\
550.831  & Fe\,{\sc i}     & $a^5\mathrm{F}-z^5\mathrm{D}^o$      &  2--3  & $-10.2~\pm~0.9$\\
552.994  & Mg\,{\sc i}     & $^1\mathrm{P}^o-^1\mathrm{D}$          &  1--2  & $-20.1~\pm~0.7$\\
553.635  & Ca\,{\sc i}     & $^3\mathrm{D}-^3\mathrm{D}^o$         &  2--1  & ~$-7.5~\pm~1.1$\\

595.834  & Fe\,{\sc i}     & $a^5\mathrm{F}-z^7\mathrm{P}^o$      &  5--4  & $-32.1~\pm~1.1$\\
613.831  & Fe\,{\sc i}     & $a^3\mathrm{H}-z^3\mathrm{G}^o$      &  4--3  & $-21.3~\pm~1.1$\\
619.327  & Fe\,{\sc i}     & $a^3\mathrm{H}-z^3\mathrm{G}^o$      &  5--4  & $-19.6~\pm~1.0$\\
624.930  & Ca\,{\sc i} ?   & $^1\mathrm{S}-^1\mathrm{P}^o$        &  0--1  & ~$-5.7~\pm~1.2$\\
625.428  & Fe\,{\sc i}     & $a^3\mathrm{H}-z^3\mathrm{G}^o$      &  6--5  & $-21.6~\pm~1.1$\\
625.809  & Fe\,{\sc i}     & $a^3\mathrm{H}-z^3\mathrm{G}^o$      &  4--4  & ~$-6.2~\pm~0.8$\\

626.687  & Fe\,{\sc i}     & $a^5\mathrm{P}-y^5\mathrm{D}^o$      &  3--3  &  $-14.0~\pm~0.9$\\
628.235  & Fe\,{\sc i}     & $a^5\mathrm{F}-z^7\mathrm{F}^o$      &  5--5  &  $-43.4~\pm~1.7$\\
631.977  & Fe\,{\sc i}     & $a^3\mathrm{H}-z^5\mathrm{G}^o$      &  4--3  &  ~$-8.0~\pm~1.0$\\
633.708  & Fe\,{\sc i}     & $a^5\mathrm{P}-y^5\mathrm{D}^o$      &  2--3  &  $-21.2~\pm~1.6$\\
636.045  & Fe\,{\sc i}     & $a^5\mathrm{F}-z^7\mathrm{F}^o$      &  5--6  &  $-58.7~\pm~1.1$\\
639.537  & Fe\,{\sc i}     & $a^3\mathrm{H}-z^5\mathrm{G}^o$      &  5--4  &  $-12.3~\pm~1.1$\\

640.209  & Fe\,{\sc i}     & $a^5\mathrm{F}-z^7\mathrm{F}^o$      &  4--4  &  $-54.0~\pm~1.3$\\
643.262  & Fe\,{\sc i}     & $a^5\mathrm{P}-y^5\mathrm{D}^o$      &  3--4  & $-21.7~\pm~1.1$\\
\noalign{\smallskip}
\hline
\end{tabular}
\end{center}
\end{table}
\setcounter{table}{1}
\begin{table}
  \caption[]{\bf (Continued)}
  \label{Table2b}
 \tabcolsep0.14cm
 \vspace*{-0.72cm}
\begin{center}
\begin{tabular}{clccc}
\hline\noalign{\smallskip}
$\lambda_{\rm lab.}$ [nm] & Ident. & Multiplet & Trans. & EW [m\AA]\\
\noalign{\smallskip}
\hline\noalign{\smallskip}
646.450  & Fe\,{\sc i}     & $a^5\mathrm{F}-z^7\mathrm{F}^o$      &  4--5  & $-22.9~\pm~1.4$ (blend)\\
646.451  & Fe\,{\sc i}     & $a^3\mathrm{H}-z^5\mathrm{G}^o$      &  4--4  & $-22.9~\pm~1.4$ (blend)\\
650.073  & Fe\,{\sc i}     & $a^5\mathrm{F}-z^7\mathrm{F}^o$      &  3--3  & $-20.1~\pm~1.0$\\
657.460  & Ca\,{\sc i}     & $^1\mathrm{S}-^3\mathrm{P}^o$        &  0--1  & $-123.9~\pm~2.1$ (blend)\\

657.604  & Fe\,{\sc i}     & $a^5\mathrm{F}-z^7\mathrm{F}^o$      &  2--2  & $-123.9~\pm~2.1$ (blend)\\
\noalign{\smallskip}
\hline
\end{tabular}
\tablefoot{For lines marked as (blend) the equivalent width given in the last column refers to the combined strength of the emission lines.}
\end{center}
\end{table}
\begin{table}
  \caption[]{Detected emission lines in the CRIRES spectra of HD\,152384.}
  \label{Table3}
 \vspace*{-0.4cm}
\begin{center}
\begin{tabular}{clccc}
\hline\noalign{\smallskip}
$\lambda_{\rm lab.}$ [nm] & Ident. & Multiplet & Trans. & EW [m\AA]\\
\noalign{\smallskip}
\hline\noalign{\smallskip}
1081.401 &  Mg\,{\sc i}    & $^3\mathrm{D}-^3\mathrm{F}^o$           &  3--4    & ~$-56.7~\pm~6.0$\\
1183.142 &  Mg\,{\sc i}    & $^1\mathrm{P}^o-^1\mathrm{S}$           &  1--0    & $-150.5~\pm~9.8$\\
1197.632 &  Fe\,{\sc i}    & $a^5\mathrm{P}-z^5\mathrm{D}^o$         &  3--4    & $-169.8~\pm~13.2$\\
1203.480 &  Si\,{\sc i}    & $^3\mathrm{P}^o-^3\mathrm{D}$           &  2--3    & $-102.5~\pm~11.7$\\
1208.697 &  Mg\,{\sc i}    & $^1\mathrm{D}-^1\mathrm{F}^o$           &  2--3    & ~$-71.8~\pm~6.6$\\
1302.039 &  Si\,{\sc i} ?  & $^1\mathrm{D}-(\frac{3}{2},\frac{3}{2})^o$  &  2--1  & ~$-56.1~\pm~14.7$\\
1488.165 &  Mg\,{\sc i}    & $^3\mathrm{D}-^3\mathrm{F}^o$           &  2--3    & $-427.4~\pm~19.8$\\
                                                            
1502.910 &  Mg\,{\sc i}    & $^3\mathrm{S}-^3\mathrm{P}^o$           &  1--2    & $-902.3~\pm~23.6$\\
1504.436 &  Mg\,{\sc i}    & $^3\mathrm{S}-^3\mathrm{P}^o$           &  1--1    & $-429.0~\pm~32.4$\\
1505.182 &  Mg\,{\sc i}    & $^3\mathrm{S}-^3\mathrm{P}^o$           &  1--0    & ~$-85.5~\pm~26.0$\\
1575.329 &  Mg\,{\sc i}    & $^3\mathrm{P}^o-^3\mathrm{D}$           &  1--2    & $-100.5~\pm~11.7$\\
1711.334 &  Mg\,{\sc i}    & $^1\mathrm{S}-^1\mathrm{P}^o$           &  0--1    & $-235.0~\pm~10.9$\\
1727.301 &  Ca\,{\sc i}    & $^1\mathrm{S}-^1\mathrm{P}^o$           &  0--1    & ~$-16.0~\pm~13.3$\\
1978.217 &  Ca\,{\sc i}    & $^3\mathrm{P}^o-^3\mathrm{D}$           &  2--3    & $-682.3~\pm~27.6$\\   
\noalign{\smallskip}
\hline
\end{tabular}
\end{center}
\end{table}
All detected emission lines in the spectra of HD\,152384 show a narrow double-peaked emission profile with a peak-to-peak separation of 103~km\,s$^{-1}$ and the blue peak appearing slightly stronger than the red peak (Figs.~\ref{Fig1} and \ref{Fig2}). The profiles of the lines due to different elements appear similar, suggesting an origin in a common circumstellar region or reservoir.

We analyse the kinematics of the gas using an average, continuum-subtracted, CRIRES line profile\footnote{For this detailed analysis of the kinematics we only use the CRIRES data, as the spectral resolution of these data is 25--35\% higher than that of the corresponding UVES spectra.}, normalised to peak intensity one, shown as the last panel in Fig.~\ref{Fig2}. We model the line profiles using a parametric flat Keplerian disk model similar to that used by \citet{vandenAncker2021}, in which the intensity of the emission decreases as $I(R) \propto (R/R_{\rm in})^{-\alpha}$, with $R$ the radial distance from the star. As in this paper, the stellar mass was taken to be 2.1 $\pm$ 0.1 M$_\odot$, and the inclination of the disk was fixed to the stellar inclination of 24$^\circ$ derived from the stellar rotation period and absorption line broadening. The model spectral line profile was convolved with a Gaussian with FWHM corresponding to the spectral resolution of the CRIRES data to simulate the observations. The inner radius $R_{\rm in}$ and outer radius $R_{\rm out}$ of the gas-emitting region were varied, as well as the power-law exponent $\alpha$. Examples of the resulting simulated spectra are shown in Fig.~\ref{Fig3}. As can be seen in this figure, the peak-to-peak separation in the line profile is sensitive to the adopted value for the outer radius $R_{\rm out}$. We find that an outer radius of $\sim$0.12 au reproduces the observed peak-to-peak separation of 103 km\,s$^{-1}$ well. Larger outer radii would produce a smaller peak-to-peak velocity separation than observed, and introduce more emission at lower absolute velocities than observed. These fitted values of $R_{\rm in}$ and $R_{\rm out}$ also depend on the adopted values of the stellar mass ($v \propto \sqrt(M_\star)$) and the disk's inclination ($v \propto \sin{i}$).

Whereas the inner radius of the gas-emitting region was largely unconstrained by the lower-resolution X-Shooter data used in \citet{vandenAncker2021}, the roughly ten times higher spectral resolution of the CRIRES data now also allows us to constrain the inner radius $R_{\rm in}$. The line profile drops off steeply at velocities smaller than -51.5 km\,s$^{-1}$, or larger than +51.5 km\,s$^{-1}$, indicating the presence of a large inner gap in the disk. We find that the depth of the central trough of the line profile, as well as the steepness of the drop-off of the wings, put tight constraints on the value of $R_{\rm in}$. Values of $R_{\rm in}$ $<$ 0.08 au produce line profiles with too strong high-velocity wings and a central trough that is too shallow to be compatible with the observed line profile. For values of $R_{\rm in}$ $>$ 0.10 au, the peaks of the line profile become too narrow, and the wings of the line profile drop off too steeply. Varying the $\alpha$ parameter, which controls the exponent in the power-law of the intensity distribution, has essentially no effect on the predicted line profile for such a small disk annulus, and is thus poorly constrained by our observations.

We conclude that the observed emission-line profiles are best explained by a narrow ring of gaseous material located at a radius of 0.09--0.12 au from the central star.
It is interesting to note that the line profiles of the different atomic species detected in our CRIRES and UVES spectra look similar. This suggests that the observed morphology of a narrow ring of gas is not due to excitation effects (where we would only see the gas at locations where the temperatures would be suitable to emit the observed lines), but instead reflect the confinement of the gas to a specific range of orbits around the star.

The fact that the blue peak in the observed line profiles appears slightly stronger than the red peak cannot be reproduced by our simple circularly symmetric model. This behaviour could be due to a slight asymmetry in the narrow ring, or there could be some additional extinction towards the red-shifted disk material. We also note that, although the exact radius of the ring is degenerate with the assumed stellar mass and inclination, the general conclusion that the observed gas-phase emission lines must originate in a ring-like geometry close to the star is independent of these assumptions.

\subsection{Excitation}
For many of the emission lines detected in HD\,152384, this is -- to the best of our knowledge -- the first identification of this specific line in any astronomical object. This points to the presence of unusual excitation conditions in the narrow ring of gas suggested by the kinematics. In this sub-section we investigate the relative strength of the lines to see what can be learned from this.

For iron and magnesium we detect a sufficient number of emission lines to analyse their intensity in a statistical way using the Self-Absorption Curve (SAC) method first developed by \cite{Friedjung1987} and subsequently applied by numerous authors \citep[e.g.][]{Muratorio1992,vandenAncker2004,Arias2006}. A practical guide to the use of the SAC is also given by \cite{Kotnik-Karuza2002}. The SAC describes the relation between the line strength and optical depth of the emission lines in a multiplet. This relation can be empirically determined by plotting $\log (F_\lambda \lambda^3/gf)$ (with $F_\lambda$ the line flux, and $gf$ the oscillator strength of the line) versus $\log gf\lambda$. For lines originating from a single multiplet, the line fluxes will fall on a single line in this diagram. This line will be horizontal for optically thin lines, whereas for optically thick emission, the points for each transition will linearly decrease with increasing $\log gf\lambda$. 

The SACs for transitions originating from different multiplets will be parallel in the diagram, but shifted with respect to each other, depending on the relative Boltzmann populations of each multiplet. Following \cite{Muratorio1988} we express these shifts as $p\chi_l$ and $q\chi_u$ for the shifts in the horizontal and vertical direction, respectively, with $\chi_l$ and $\chi_u$ the excitation potentials of the lower and upper terms in the transitions, normalised to a reference Boltzmann temperature of 5040~K (as also explained in \cite{Kotnik-Karuza2002}). Although in principle in the SAC method the excitation temperatures for the lower and upper energy levels can be different, we assume in our analysis of HD\,152384 that the detected gas is in local thermodynamic equilibrium, and the excitation temperatures of the upper and lower levels are the same, i.e. $p = q$.

\begin{figure}[t]
   \centering
   \includegraphics[width=9.0cm]{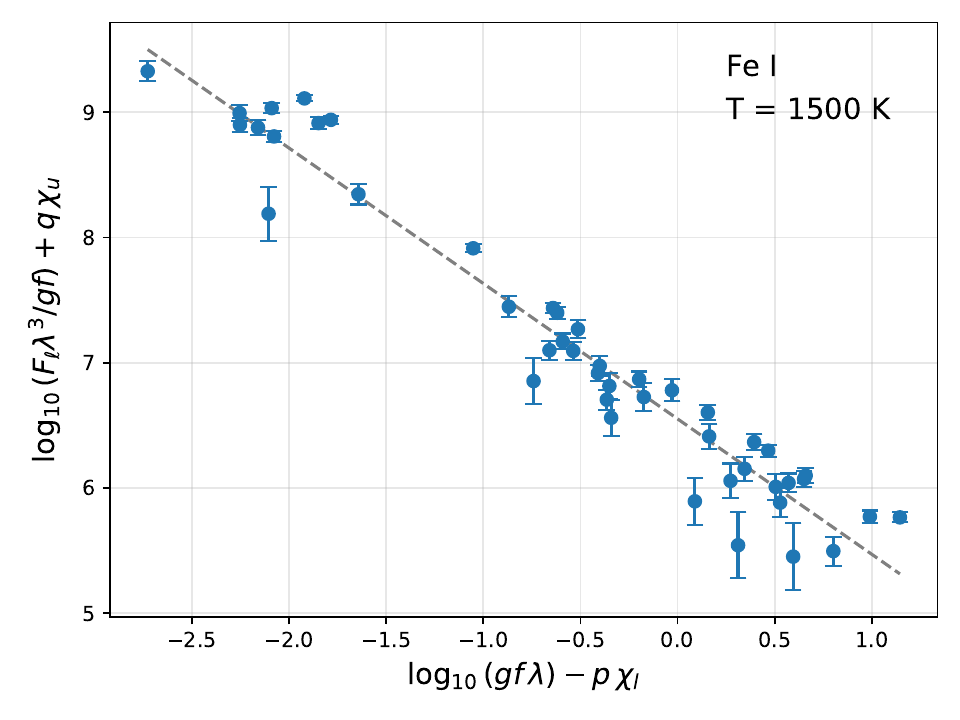}
   \caption{Emission-line Self-Absorption Curve (SAC) for the Fe\,{\sc i} emission lines detected in HD\,152384. The dashed grey line shows the linear fit to the data points used to derive upper limits to the column density and size of the emitting region.}
              \label{Fig4}
    \end{figure}

\begin{figure}[t]
   \centering
   \includegraphics[width=9.0cm]{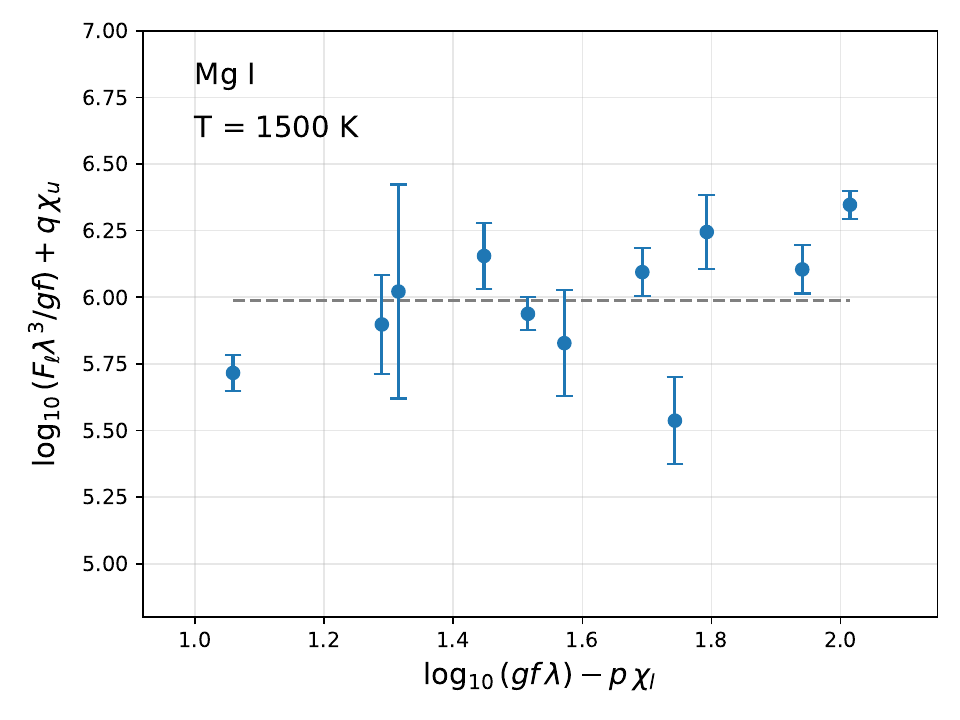}
   \caption{Emission-line Self-Absorption Curve (SAC) for the Mg\,{\sc i} emission lines detected in HD\,152384. The dashed grey line shows the fit to the data points used to derive the column density and size of the emitting region.}
              \label{Fig5}
    \end{figure}

Using oscillator strengths $\log gf$ and energy levels from the atomic database by \cite{vanHoof2018}\footnote{\url{https://linelist.pa.uky.edu/newpage/}}, we create empirical SACs for the Fe\,{\sc i} and Mg\,{\sc i} lines detected in our UVES and CRIRES spectra of HD\,152384. When constructing the SACs for HD\,152384, the line fluxes $F_\lambda$ were computed from the measured equivalent widths given in Tables~\ref{Table2} and \ref{Table3} and the continuum fluxes at the wavelength of the line from the photospheric model of HD\,152384 shown in \cite{vandenAncker2021}. 
From this analysis we exclude the transitions indicated as ``blend'' in Table~\ref{Table2}, as no unambiguous line flux could be determined for these lines. For constructing our SACs we find that when assuming a Boltzmann temperature for both the upper and lower energy levels of around 1500~K, the lines from the various detected multiplets form a single line in the diagrams (Figs.~\ref{Fig4} and \ref{Fig5}). As can be seen from Fig.~\ref{Fig4}, all Fe\,{\sc i} line fluxes are located on the decreasing part of the SAC, indicating that all detected Fe\,{\sc i} lines are at least partly optically thick. In contrast to this, the Mg\,{\sc i} lines form a horizontal line (Fig.~\ref{Fig5}), indicating that they are mostly optically thin.

In the SAC, the horizontal- and vertical-axis intercepts in the diagram, $X_0$ and $Y_0$, provide a measure for the column density and spatial extent of the line-emitting region. For spectra with only optically thick lines, one can only obtain the decreasing part of the SAC curve, so that the method then only gives a lower limit to the column density and an upper limit to the extension of the emission-line region. Fitting a straight line to the observed data points in Fig.~\ref{Fig4}, we obtain the values of the x- and y-axis intersections of the SAC, $X_0$ and $Y_0$. Using these and equations (3) and (4) given by \cite{Muratorio2006} we obtain a lower limit to the column density of the Fe\,{\sc i} emission of 3.4 $\times$ 10$^{23}$~cm$^{-2}$, and an upper limit to the extent of the Fe\,{\sc i} emitting region of 1.2 $\times$ 10$^{25}$~cm$^{2}$. We note that for optically thick lines the column density and the extent of the emitting region may be degenerate, so caution should be exercised when combining these values. 
For the optically thin Mg\,{\sc i} lines we obtain a column density of 2.2 $\times$ 10$^{22}$~cm$^{-2}$, and an extent of the emitting region of 2.4 $\times$ 10$^{25}$~cm$^{2}$. Combining this column density and emitting region, we can then obtain an estimate of the mass of the emitting Mg\,{\sc i} gas, which is 2.2 $\times$ 10$^{22}$~kg, or 3.6 $\times$ 10$^{-3}$~M$_\oplus$. We stress that due to the lack of detail in our modelling on e.g. optical depth, or the implicit assumptions on geometry in the SAC analysis, this derived mass is quite uncertain. However, it may still serve as a useful guide to the order of magnitude of material required to explain the presence of the detected emission lines. 

\section{Discussion}
        \cite{Kospal2013} found the debris disk around the 30 Myr old star HD\,21997 to be unexpectedly CO-rich. Subsequently the term hybrid disk was coined to indicate such systems in which the dust is likely of second generation, but the gas potentially primordial. \citet{Wyatt2015} introduced the 12~$\mu$m excess as another empirical criterion for identifying disks caught in the transition stage between protoplanetary and debris disks. This method was used by \citet{Iglesias2023} to identify hybrid disk candidates, one of which is HD\,152384. However, many CO-rich disks only have cold exoKuiper belts, so they do not exhibit a strong 12~$\mu$m excess. These systems will thus not be included in the samples of \cite{Wyatt2015} and \cite{Iglesias2023}.

We note that some hybrid disks also show gas emission in tracers such as H$\alpha$ or the calcium lines -- for example in the prototypical hybrid disk \object{HD\,141569} \citep{Thi2014,Miley2018,DiFolco2020}. This atomic gas is likely located much closer to the central star than the cold gas seen in CO lines, and could be primordial in origin \citep{Kospal2013,Nakatani2021,Nakatani2023}. 
The group of ``hybrid'' disks may partly overlap with the population of so-called ``extreme'' debris disks, which typically exhibit high fractional dust luminosities (dust fractional luminosity, $L_{\rm d}/L_\star$ $\ge$ 10$^{-2}$) due to the presence of warm dust. Most (but not all) extreme debris disks also exhibit significant variability in their infrared flux -- presumably due to dust production by unusually high rates of collisional activity \citep[e.g.][]{Meng2014,Su2019,Rieke2021,Moor2021,Moor2022}. However, whereas the group of ``hybrid'' disks is limited to relatively young stars (ages $<$ 40 Myr), extreme debris disk phenomena may also appear in much older systems.

HD\,152384 shows evidence for both circumstellar dust (through the presence of an infrared excess in its SED; \citealt{vandenAncker2021,Iglesias2023}), and gas, through the gaseous atomic emission lines reported here. The morphology of the dust around HD\,152384 is unknown. However, the peak of the infrared excess in this star is located around 10~$\mu$m \citep{vandenAncker2021}, suggesting the presence of small dust grains of temperatures of a few hundred Kelvin. For dust in blackbody equilibrium with the radiation field of the central star, this dust would be located at typical radii of $\sim$1--5 au from the central star. In contrast to the dust, the circumstellar atomic gas in HD\,152384 appears to be confined to a narrow ring at an inferred location of only $\sim$0.1 au. Interestingly, we note that the dust equilibrium temperature at 0.1 au of an A0\,V star is 1500--1600~K, i.e. close to the silicate sublimation temperature. The main species observed in our data, iron, magnesium, calcium and silicon, are consistent with an origin in small olivine or pyroxene dust grains, similar to the composition of dust observed in many protoplanetary disks \citep[e.g.][]{Waters2015,Birnstiel2024}.

The phenomenology in HD\,152384 of atomic gas in the inner disk, combined with dust at several au is similar to what has been observed with JWST in HD\,172555 \citep{Samland2025}. In this system, the presence of a close-in source of gas could be linked to the dynamical processes associated with the giant impact that is proposed to explain the warm dust belt. It is plausible that similar processes could have played a role in creating the atomic gas seen in HD\,152384. 
The collisional scenario presented above could also offer a natural explanation for the relatively high ($L_{\rm IR}/L_\star$ = 0.05; \citealt{vandenAncker2021}) fractional infrared excess of HD\,152384, suggesting the existence of a much larger amount of small, warm dust grains than is typical in debris disk systems. We note that HD\,152384 is one of the youngest gas-bearing debris disk systems identified to date and the peculiar characteristics of this enigmatic system could thus very well be a reflection of a much more violent early phase of the evolution of debris disks in general.

Our derived HD\,152384 Mg\,{\sc i} mass of 2.2 $\times$ 10$^{22}$~kg is compatible with an origin in a planetesimal-mass parent body. Assuming a typical dust composition of olivines and pyroxenes, with 50\% iron-rich and 50\% Mg-rich dust (compatible with the upper limits derived from the optically thick Fe\,{\sc i} lines), the mass of the parent dust population responsible for the observed gaseous emission would be 6.3 $\times$ 10$^{22}$~kg or about 0.01~M$_\oplus$. This amount of small dust can easily be generated in the collisional cascade following the collision of one or more large planetesimals \citep[e.g.][]{Wyatt2007,Gaspar2013,MacGregor2016}.

Both the location close to the dust sublimation radius, as well as the strong emission from refractory species point towards sublimation of dust grains as the source of the circumstellar gas seen in our UVES and CRIRES spectra. Given the absence of other tracers of disk activity in HD\,152384, we postulate that the most likely source of the dust particles which are the source of the material in the narrow gaseous ring are collisions among planetesimals or protoplanets. These collisions could be caused by dynamic clearing due to planet migration, or stirring of the planetesimal belt due to the presence of a wide binary companion in an eccentric orbit. The collisions would likely be at locations further away from the central star, but the small (1~$\mu$m--1~mm) dust grains produced in these collisions would drift inwards due to Poynting-Robertson drag. The strong radiation pressure around an early A-type star is expected to counteract this inward motion and remove the small dust grains on short time scales. However, in the presence of a gas disk this picture may change and gas can trap grains larger than the blowout size exterior to the sublimation region \citep[e.g.][]{Pearce2020}. Interestingly, simulations for the transport of mm-sized dust grains around an A0 star show final distances for mm-sized grains to be just beyond 0.1 au in the presence of even moderately dense gas at the sublimation radius \citep{Pearce2020}. 

For the scenario noted above of stirring of a planetary belt by a binary companion, we note that the current position of the wide (projected separation of 205 au) companion to HD\,152384 detected in the Gaia and VLT/ISAAC data is likely too far away from the A0\,V star to cause significant perturbations in a dust belt located at 1--5~au from the central star. However, if this companion is on a highly eccentric orbit, it could perturb the dust belt during periastron passage. Alternatively, the potentially much closer-in companion hinted at by the proper-motion anomaly reported by \cite{Kervella2022} could also be a viable source of such perturbations. Further high angular resolution observations of HD\,152384 would be useful to place more stringent limits on the existence of such a perturber than is possible with the current data.

The observed phenomenology of the presence of atomic circumstellar gas of refractory elements in a system which no longer shows accretion onto the central star makes the HD\,152384 system unique in the group of young stars. However, there are systems which show some similarity to HD\,152384. For example, 
\cite{Thanathibodee2026} report evidence for the presence of variable thermal emission due to warm (1,400\,K), optically thick, large ($\ge$\,1~$\mu$m) dust grains near the dust sublimation radius in the transitional disk around the T~Tauri star \object{CVSO\,1942}. These authors propose that the rapid dust appearance seen in this object may indicate in situ dust formation, possibly from planetesimal collisions in the inner disk. \cite{Szewczyk2026} report detections of both CO and atomic Ca\,{\sc ii} gas in the circumstellar environment of the 2.1~Myr old debris disk-bearing intermediate-mass star \object{HD\,44892}, which these authors take as indications for a late gas dispersal stage of a protoplanetary disk. We note that so far no CO detection has been reported in either HD\,152384 or CVSO\,1942, complicating the comparison of these systems with true hybrid disks.

Metallic gas and volatiles have also been detected in other debris disk systems. \object{HD\,172555} may be the clearest case of a rocky, metal-rich circumstellar environment, where the presence of silicate dust, SiO gas, and atomic oxygen has been interpreted as the result of a high-energy impact among differentiated bodies rather than a volatile-rich exocometary reservoir \citep[e.g.][]{Lisse2009,Samland2025,Roumeliotis2026}. In contrast, $\beta$~Pic hosts one of the best-studied gas-bearing debris disks, with a substantial volatile component dominated by C, O, and N. Its gas appears to be overabundant in carbon relative to solar composition, consistent with secondary production from icy planetesimals and/or preferential retention of low-radiation-pressure species \citep{Roberge2006,Brandeker2016,Cataldi2018}.

The stars \object{HD\,110058} and \object{HD\,36546} both show cold CO-rich gas detected with ALMA, indicating the presence of volatile-bearing debris disks in which the molecular gas is most naturally explained as secondary, exocometary material released from colliding planetesimals \citep{Hales2022,Rebollido2022}. The \object{HD\,131488} system is likewise notable for CO-bearing gas in the terrestrial-planet region, making it an additional example of a volatile-rich debris disk rather than a purely metal-gas system \citep[e.g.][]{Pawellek2024,Lu2026}. Thermochemical models for the detected CO gas favour a secondary origin for the gas in these disks \citep{Hales2019}.

Although HD\,152384 appears to be unique within the group of young stars, there is another class of stars that show a surprisingly similar phenomenology: metal-polluted white dwarfs \citep[e.g.][]{Gansicke2006,Wilson2014,Manser2016,Melis2020,GentileFusillo2021,Veras2024} form a rare class of emission-line systems which exhibit infrared excess in their SEDs from circumstellar dust. In these systems, this dust is believed to originate from the tidal disruption of rocky planetesimals, while the characteristic double-peaked atomic emission lines seen in these systems -- strikingly similar to those observed in HD\,152384 -- trace a gaseous component within the debris disk. The gas production mechanism remains uncertain, with proposed origins including dust sublimation near the inner disk edge \citep{Metzger2012}, collisional cascades \citep{Kenyon2017}, or perturbations in the disk from an embedded planetesimal \citep{Manser2019}.

The composition of the circumstellar gas in HD\,152384 also shows some similarity to other well-studied debris disk systems such as \object{$\beta$ Pic}. In these systems, the dust originates primarily from the collisional grinding of unseen planetesimal belts, which undergo a steady-state collisional cascade converting kilometre-sized bodies into micron-sized grains rapidly removed by radiation pressure, stellar winds, and Poynting-Robertson drag \citep[e.g.][]{Wyatt2008,Krivov2010,Iglesias2025}. In the $\beta$ Pictoris system specifically, modelling and spectroscopy reveal contributions from both destructive collisions among large planetesimals at tens to hundreds of au and from sublimating exocomets on star-grazing orbits, injecting fresh, porous material \citep{Lecavelier1998,Matra2019}. Transient clumps and asymmetries further indicate episodic boosts from recent catastrophic events, such as planetesimal disruptions perturbed by planets \citep[e.g.][]{Han2023}. Many debris disks also display ring-like morphologies (e.g. \object{HD\,107146}, \citealt{Marino2018,Marino2026}; \object{Fomalhaut}, \citealt{MacGregor2017}). The asymmetry seen in the red- and blue-shifted line profiles seen in HD\,152384 could also be due to clumps similar to those seen in both dust and gas in the $\beta$ Pic disk \citep[e.g.][]{Dent2014}. We note, however, that the atomic emission lines seen in HD\,152384 originate much closer to the central star than the clumps seen in $\beta$ Pic, so although the phenomenology may be similar, the underlying physical mechanism for producing the asymmetry may very well be different.

\section{Conclusions}
High-resolution UVES and CRIRES spectroscopy of HD\,152384 reveals a rich spectrum of 74 narrow, double-peaked emission lines from the refractory elements iron, magnesium, silicon and calcium. The line profiles, with 103~km\,s$^{-1}$ peak separation, trace stable Keplerian motion in a narrow ring at $R \simeq$ 0.09--0.12~au, located precisely at the silicate dust sublimation radius (1500--1600~K) of this young A0\,V star.

Although atomic gas is now commonly detected in debris disks \citep[e.g.][]{Hales2017,Rebollido2018,Iglesias2025}, these detections are mostly limited to -- sometimes transient -- {\it absorption} in lines such as the Ca\,{\sc ii} H and K lines and the Na\,{\sc i} D doublet. To the best of our knowledge, the phenomenology reported here for HD\,152384 -- with {\it emission} from a large number of gaseous neutral atomic lines -- has not been previously reported within the group of debris disks. However, we do note that similar emission lines are seen in metal-polluted white dwarfs \citep{Gansicke2006,Manser2016}, where secondary gas production is also believed to proceed via sublimation of collisionally produced dust. There are also some debris disk systems such as HD\,172555 where molecular gas is found close to the star. We speculate that similar phenomena could be observed in other young debris disk systems with high levels of collisional activity, and a dedicated search for such activity would be a fruitful avenue to characterise the circumstellar material in debris disks.

We conclude that the geometry and refractory-rich gas composition mark HD\,152384 as a prime example of a transitional object between primordial protoplanetary gas disks and mature debris disks. In this system, the circumstellar gas originates from thermal or collisional processing of refractory dust grains in the innermost disk, likely triggered by dynamical clearing from planet migration or wide binary interactions. HD\,152384 thus represents one of the youngest known examples of secondary gas production in a debris disk system, offering critical insights into the violent early phases of rocky planetesimal evolution around intermediate-mass stars.

\section*{Data availability}
The reduced UVES and CRIRES spectra used in this article are available as advanced data products in the ESO science archive facility: \url{https://archive.eso.org/wdb/wdb/adp/phase3_spectral/query}
\\

\begin{acknowledgements}
DPI and OP acknowledge support from the Science and Technology Facilities Council via grant number ST/X001016/1. 
We thank the ESO staff at Paranal and in Garching for their support during the preparation and execution of the observations. 
We would also like to thank the anonymous referee for providing insightful comments which helped to substantially improve the manuscript. 
This work made use of Astropy\footnote{\url{https://www.astropy.org}}: a community-developed core Python package and an ecosystem of tools and resources for astronomy \citep{Astropy2013, Astropy2018, Astropy2022}.
\end{acknowledgements}

 \bibliographystyle{apa}
 \bibliography{bibliography}

\begin{thebibliography}{}

\bibitem[\protect\citeauthoryear{Astropy Collaboration et al.}{2013}]{Astropy2013} Astropy Collaboration, Robitaille, T.~P., Tollerud, E.~J., Greenfield, P., Droettboom, M., Bray, E., Aldcroft, T., et al. 2013, A\&A, 558, A33.

\bibitem[\protect\citeauthoryear{Astropy Collaboration et al.}{2018}]{Astropy2018} Astropy Collaboration, Price-Whelan, A.~M., Sip{\H{o}}cz, B.~M., G{\"u}nther, H.~M., Lim, P.~L., Crawford, S.~M., Conseil, S., et al. 2018, AJ, 156, 123

\bibitem[\protect\citeauthoryear{Astropy Collaboration et al.}{2022}]{Astropy2022} Astropy Collaboration, Price-Whelan, A.~M., Lim, P.~L., Earl, N., Starkman, N., Bradley, L., Shupe, D.~L., et al. 2022, ApJ, 935, 167.
  
\bibitem[\protect\astroncite{Arias et~al.}{2006}]{Arias2006} Arias, M.~L., Zorec, J., Cidale, L., Ringuelet, A.~E., Morell, N.~I., Ballereau, D., 2006 A\&A, 460, 821

\bibitem[\protect\astroncite{Birnstiel}{2024}]{Birnstiel2024} Birnstiel, T. 2024, ARA\&A, 62, 157

\bibitem[\protect\astroncite{Brandeker et~al.}{2016}]{Brandeker2016} Brandeker, A., Cataldi, G., Olofsson, G., Vandenbussche, B., Acke, B., Barlow, M.~J., Blommaert, J.~A.~D.~L., et al. 2016, A\&A, 591, A27

\bibitem[\protect\astroncite{Cataldi et~al.}{2018}]{Cataldi2018} Cataldi, G., Brandeker, A., Wu, Y., Chen, C., Dent, W., de Vries, B.~L., Kamp, I., et al. 2018, ApJ, 861, 72

\bibitem[\protect\astroncite{Dekker et~al.}{2000}]{Dekker2000} Dekker, H., D’Odorico, S., Kaufer, A., Delabre, B., Kotzlowski, H. 2000, in
Iye, M., Moorwood, A.~F., eds, Proc. SPIE, Vol. 4008, Optical and IR Telescope Instrumentation and Detectors. pp 534–-545

\bibitem[\protect\astroncite{Dent et~al.}{2014}]{Dent2014} Dent, W.~R.~F., Wyatt, M.~C., Roberge, A., et al. 2014, Science, 343, 1490

\bibitem[\protect\astroncite{Di Folco et~al.}{2020}]{DiFolco2020} Di Folco, E., P\'ericaud, J., Dutrey, A., Augereau, J.~C., Chapillon, E., Guilloteau, S., Pi\'etu, V., Boccaletti, A. 2020, A\&A, 635, A94
  
\bibitem[\protect\astroncite{Dorn et~al.}{2023}]{Dorn2023} Dorn, R.~J., Bristow, P., Smoker, J.~V., et al. 2021, A\&A, 671, A24
  
\bibitem[\protect\astroncite{Friedjung \& Muratorio}{1987}]{Friedjung1987} Friedjung, M., Muratorio, G. 1987, A\&A, 188, 100

\bibitem[\protect\astroncite{Gaia Collaboration}{2023}]{Gaia2023} Gaia Collaboration, et al. 2023, A\&A, 674, A1

\bibitem[\protect\astroncite{G\"ansicke et~al.}{2006}]{Gansicke2006} G{\"a}nsicke, B.~T., Marsh, T.~R., Southworth, J., Rebassa-Mansergas, A. 2006, Science, 314, 1908

\bibitem[\protect\astroncite{G{\'a}sp{\'a}r et~al.}{2013}]{Gaspar2013} G{\'a}sp{\'a}r, A., Rieke, G.~H., Balog, Z. 2013, ApJ, 768, 25
 
\bibitem[\protect\astroncite{Gentile Fusillo et~al.}{2021}]{GentileFusillo2021} Gentile Fusillo, N.~P., Manser, C.~J., G{\"a}nsicke, B.~T., Toloza, O., Koester, D., Dennihy, E., Brown, W.~R., et al. 2021, MNRAS, 504, 2707
   
\bibitem[\protect\astroncite{Hales et~al.}{2017}]{Hales2017} Hales, A.~S., Barlow, M.~J., Crawford, I.~A., et al. 2017 MNRAS, 466, 3582

\bibitem[\protect\astroncite{Hales et~al.}{2019}]{Hales2019} Hales, A.~S., Gorti, U., Carpenter, J.~M., Hughes, M., Flaherty, K. 2019, ApJ, 878, 113

\bibitem[\protect\astroncite{Hales et~al.}{2022}]{Hales2022} Hales, A.~S., Marino, S., Sheehan, P.~D., Ulloa, S., P{\'e}rez, S., Matr{\`a}, L., Kral, Q., et al. 2022, ApJ, 940, 161

\bibitem[\protect\astroncite{Han et~al.}{2023}]{Han2023} Han, Y., Wyatt, M.~C., Dent, W.~R.~F. 2023, MNRAS, 519, 3257

\bibitem[\protect\astroncite{Hughes et~al.}{2018}]{Hughes2018} Hughes, A.~M., Duch{\^e}ne, G., Matthews, B.~C. 2018, ARA\&A, 56, 541

\bibitem[\protect\astroncite{Iglesias et~al.}{2023}]{Iglesias2023} Iglesias, D.~P., Pani\'c, O., van den Ancker, M.~E., Petr-Gotzens, M.~G., Siess, L., Vioque, M., Pascucci, I., Oudmaijer, R.~D., Miley, J. 2023, MNRAS, 519, 3958

\bibitem[\protect\astroncite{Iglesias et~al.}{2025}]{Iglesias2025} Iglesias, D., Rebollido, I., Norazman, A., et al. 2025, Space Science Reviews, 221, 8, 122

\bibitem[\protect\astroncite{Kenyon \& Bromley}{2017}]{Kenyon2017} Kenyon, S.~J., Bromley, B.~C. 2017, ApJ, 844, 116

\bibitem[\protect\astroncite{Kervella et~al.}{2022}]{Kervella2022} Kervella, P., Arenou, F., Th\'evenin. F. 2022, A\&A, 657, A7

\bibitem[\protect\astroncite{Kiefer et~al.}{2014}]{Kiefer2014} Kiefer, F., Lecavelier des Etangs, A., Boissier, J., Vidal-Madjar, A., Beust, H., Lagrange, A.-M., H{\'e}brard, G., et al. 2014, Nature, 514, 462

\bibitem[\protect\astroncite{K{\'o}sp{\'a}l et~al.}{2013}]{Kospal2013} K{\'o}sp{\'a}l, {\'A}., Mo{\'o}r, A., Juh{\'a}sz, A., {\'A}brah{\'a}m, P., Apai, D., Csengeri, T., Grady, C.~A. et al., 2013, ApJ, 776, 77

\bibitem[\protect\astroncite{Kotnik-Karuza et~al.}{2002}]{Kotnik-Karuza2002} Kotnik-Karuza, D., Friedjung, M., Selvelli, P.~L. 2002, Ap\&SS, 280, 413

\bibitem[\protect\astroncite{Krivov}{2010}]{Krivov2010} Krivov, A.~V. 2010, Research in Astronomy and Astrophysics, 10, 383
  
\bibitem[\protect\astroncite{Lecavelier des Etangs}{1998}]{Lecavelier1998} Lecavelier Des Etangs, A. 1998, A\&A, 337, 501

\bibitem[\protect\astroncite{Lisse}{2009}]{Lisse2009} Lisse, C.~M., Chen, C.~H., Wyatt, M.~C., Morlok, A., Song, I., Bryden, G., Sheehan, P. 2009, ApJ, 701, 2019

\bibitem[\protect\astroncite{Lu}{2009}]{Lu2026} Lu, C.~X., Rebollido, I., Brittain, S., Beck, T., Chen, C.~H., Worthen, K., Najita, J., et al. 2026, ApJ, 998, 40
  
\bibitem[\protect\astroncite{MacGregor et~al.}{2016}]{MacGregor2016} MacGregor, M.~A., Wilner, D.~J., Chandler, C., Ricci, L., Maddison, S.~T., Cranmer, S.~R., Andrews, S.~M. et al., 2016, ApJ, 823, 79

\bibitem[\protect\astroncite{MacGregor et~al.}{2017}]{MacGregor2017} MacGregor, M.~A., Matr{\`a}, L., Kalas, P., et al. 2017, ApJ, 842, 8

\bibitem[\protect\astroncite{Manser et~al.}{2016}]{Manser2016} Manser, C.~J., G{\"a}nsicke, B.~T., Marsh, T.~R., Veras, D., Koester, D., Breedt, E., Pala, A.~F., et al. 2016, MNRAS, 455, 4467

\bibitem[\protect\astroncite{Manser et~al.}{2019}]{Manser2019} Manser, C.~J., G{\"a}nsicke, B.~T., Eggl, S., Hollands, M., Izquierdo, P., Koester, D., Landstreet, J.~D. et al. 2019, Science, 364, 66

\bibitem[\protect\astroncite{Marino et~al.}{2018}]{Marino2018} Marino, S., Carpenter, J., Wyatt, M.~C., et al. 2018, MNRAS, 479, 5423

\bibitem[\protect\astroncite{Marino et~al.}{2020}]{Marino2020} Marino, S., Flock, M., Henning, T., Kral, Q., Matr{\`a}, L., Wyatt, M.~C. 2020, MNRAS, 492, 4409
  
\bibitem[\protect\astroncite{Marino et~al.}{2026}]{Marino2026} Marino, S., Matr{\'a}, L., Hughes, A.~M., Erhardt, J., et al. 2026 A\&A, 705, A195
  
\bibitem[\protect\astroncite{Matr{\'a} et~al.}{2019}]{Matra2019} Matr{\`a}, L., Wyatt, M.~C., Wilner, D.~J., et al. 2019, AJ, 157, 135
  
\bibitem[\protect\astroncite{Melis et~al.}{2020}]{Melis2020} Melis, C., Klein, B., Doyle, A.~E., Weinberger, A., Zuckerman, B., Dufour, P. 2020, ApJ, 905, 56
  
\bibitem[\protect\astroncite{Meng et~al.}{2014}]{Meng2014} Meng, H.~Y.ZA., Su, K.~Y.~L., Rieke, G.~H., et al., 2014 Science, 345, 1032
  
\bibitem[\protect\astroncite{Metzger et~al.}{2012}]{Metzger2012} Metzger, B.~D., Rafikov, R.~R., Bochkarev, K.~V. 2012, MNRAS, 423, 505

\bibitem[\protect\astroncite{Miley et~al.}{2018}]{Miley2018} Miley, J.~M., Pani\'c, O., Wyatt, M., Kennedy, G.~M. 2018, A\&A, 615, L10
  
\bibitem[\protect\astroncite{Mo\'or et~al.}{2021}]{Moor2021} Mo\'or, A., \'Abrah\'am, P., Szab\'o, G., Vida, K., et al. 2021, ApJ, 910, 27

\bibitem[\protect\astroncite{Mo\'or et~al.}{2022}]{Moor2022} Mo{\'o}r, A., {\'A}brah{\'a}m, P., K{\'o}sp{\'a}l, {\'A}., Su, K.~Y.~L., Rieke, G.~H., Vida, K., Cataldi, G., et al. 2022, MNRAS, 516, 5684
  
\bibitem[\protect\astroncite{Moorwood et~al.}{1998}]{Moorwood1998} Moorwood, A., et al. 1998, The Messenger, 94, 7

\bibitem[\protect\astroncite{Muratorio \& Friedjung}{1988}]{Muratorio1988} Muratorio, G., Friedjung, M. 1988, A\&A, 190, 103
  
\bibitem[\protect\astroncite{Muratorio et~al.}{1992}]{Muratorio1992} Muratorio, G., Viotti, R., Friedjung, M., Baratta, G.~B., Rossi, C. 1992, A\&A, 258, 423

\bibitem[\protect\astroncite{Muratorio et~al.}{2006}]{Muratorio2006} Muratorio, G., Rossi, C., Friedjung, M. 2006, A\&A, 450, 593 
  
\bibitem[\protect\astroncite{Nakatani et~al.}{2021}]{Nakatani2021} Nakatani, R., Kobayashi, H., Kuiper, R., Nomura, H., Aikawa, Y. 2021, ApJ, 915, 90

\bibitem[\protect\astroncite{Nakatani et~al.}{2023}]{Nakatani2023} Nakatani, R., Turner, N.~J., Hasegawa, Y., Cataldi, G., Aikawa, Y., Marino, S., Kobayashi, H. 2023, ApJL, 959, L28

\bibitem[\protect\astroncite{Pawellek et~al.}{2024}]{Pawellek2024} Pawellek, N., Mo{\'o}r, A., Kirchschlager, F., Milli, J., K{\'o}sp{\'a}l, {\'A}., {\'A}brah{\'a}m, P., Marino, S., et al. 2024, MNRAS, 527, 3559
    
\bibitem[\protect\astroncite{Pearce et~al.}{2020}]{Pearce2020} Pearce, T.~D., Krivov, A.~V., Booth, M. 2020, MNRAS, 498, 2798
  
\bibitem[\protect\astroncite{Rebollido et~al.}{2018}]{Rebollido2018} Rebollido, I., Eiroa, C., Montesinos, B., et al. 2018, A\&A, 614, A3

\bibitem[\protect\astroncite{Rebollido et~al.}{2022}]{Rebollido2022} Rebollido, I., Ribas, {\'A}., de Gregorio-Monsalvo, I., Villaver, E., Montesinos, B., Chen, C., Canovas, H., et al. 2022, MNRAS, 509, 693

\bibitem[\protect\astroncite{Rieke et~al.}{2021}]{Rieke2021} Rieke, G.~H., Su, K.~Y.~L., Melis, C., G\'asp\'ar, A. 2021, ApJ, 918, 71

\bibitem[\protect\astroncite{Roberge et~al.}{2006}]{Roberge2006} Roberge, A., Feldman, P.~D., Weinberger, A.~J., Deleuil, M., Bouret, J.-C. 2006, Natur, 441, 724

\bibitem[\protect\astroncite{Roumeliotis et~al.}{2026}]{Roumeliotis2026} Roumeliotis, Z., Matr{\`a}, L., Kennedy, G.~M., Marino, S., Su, K.~Y.~L., Wilner, D.~J., Wyatt, M.~C., et al., 2026, A\&A, 706, A4
  
\bibitem[\protect\astroncite{Samland et~al.}{2025}]{Samland2025} Samland, M., Henning, T., Caratti o Garatti, A., Giannini, T., Bouwman, J., Tabone, B., Arabhavi, A.~M., et al., 2025, ApJ, 989, 132
  
\bibitem[\protect\astroncite{Szewczyk et~al.}{2026}]{Szewczyk2026} Szewczyk, K.~M., Pani\'c, O., Iglesias, D.~P., Pearce, T.~D., Miley, J.~M. 2026, MNRAS, 545, 2150

\bibitem[\protect\astroncite{Su et~al.}{2019}]{Su2019} Su, K.~Y.~L., Jackson, A.~P., G\'asp\'ar, A., Rieke, G.~H., et al. 2019, AJ, 157, 202
  
\bibitem[\protect\astroncite{Thanathibodee et~al.}{2026}]{Thanathibodee2026} Thanathibodee, T., Espaillat, C., Calvet, N., et al. 2026, ApJL, 999, L31
  
\bibitem[\protect\astroncite{Thi et~al.}{2014}]{Thi2014} Thi, W.-F., Pinte, C., Pantin, E., et al. 2014, A\&A, 561, A50
  
\bibitem[\protect\astroncite{van den Ancker et~al.}{2004}]{vandenAncker2004} van den Ancker, M.~E., Blondel, P.~F.~C., Tjin A Djie, H.~R.~E., Grankin, O.~V., Ezhkova, V., Shevchenko, S., et al. 2004, MNRAS, 349, 1516

\bibitem[\protect\astroncite{van den Ancker et~al.}{2021}]{vandenAncker2021} van den Ancker, M.~E., Gentile Fusillo, N.~P., Haworth, T.~J., Manara, C.~F., Miles-P\'aez, P.~A., Oudmaijer, R.~D., Pani\'c, O., Petit dit de la Roche, D.~J.~M., Petr-Gotzens, M.~G., Vioque, M. 2021, A\&A, 651, L11

\bibitem[\protect\astroncite{van der Marel}{2023}]{vanderMarel2023} van der Marel, N., 2023, European Physical Journal Plus, 138, 225

\bibitem[\protect\astroncite{van Hoof}{2018}]{vanHoof2018} van Hoof, P.~A.~M. 2018, Galaxies, 6, 63

\bibitem[\protect\astroncite{Veras et~al.}{2024}]{Veras2024} Veras, D., Mustill, A.~J., Bonsor, A. 2024, Reviews in Mineralogy and Geochemistry, 90, 141

\bibitem[\protect\astroncite{Waters}{2015}]{Waters2015} Waters, L.~B.~F.~M. 2015, European Physical Journal Web of Conferences, 102, 3

\bibitem[\protect\astroncite{Wilson et~al.}{2014}]{Wilson2014} Wilson, D.~J., G{\"a}nsicke, B.~T., Koester, D., Raddi, R., Breedt, E., Southworth, J., Parsons, S.~G. 2014, MNRAS, 445, 1878

\bibitem[\protect\astroncite{Wyatt}{2008}]{Wyatt2008} Wyatt, M.~C. 2008, ARA\&A 46, 339

\bibitem[\protect\astroncite{Wyatt et~al.}{2007}]{Wyatt2007} Wyatt, M.~C., Smith, R., Greaves, J.~S., Beichman, C.~A., Bryden, G., Lisse, C.~M. 2007, ApJ, 658, 569

\bibitem[\protect\astroncite{Wyatt et~al.}{2015}]{Wyatt2015} Wyatt, M.~C., Pani{\'c} O., Kennedy G.~M., Matr{\`a} L. 2015, Ap\&SS, 357, 103
%
\end{thebibliography}

\end{document}